\documentstyle[epsf]{mn}
\newif\ifAMStwofonts
\AMStwofontstrue

\ifoldfss
  \ifCUPmtlplainloaded \else
    \NewTextAlphabet{textbfit} {cmbxti10} {}
    \NewTextAlphabet{textbfss} {cmssbx10} {}
    \NewMathAlphabet{mathbfit} {cmbxti10} {} 
    \NewMathAlphabet{mathbfss} {cmssbx10} {} 
  \fi
  \ifAMStwofonts
    \ifCUPmtlplainloaded \else
      \NewSymbolFont{upmath} {eurm10}
      \NewSymbolFont{AMSa} {msam10}
      \NewMathSymbol{\upi}     {0}{upmath}{19}
      \NewMathSymbol{\umu}     {0}{upmath}{16}
      \NewMathSymbol{\upartial}{0}{upmath}{40}
      \NewMathSymbol{\leqslant}{3}{AMSa}{36}
      \NewMathSymbol{\geqslant}{3}{AMSa}{3E}

    \fi
  \fi
\fi 

\ifnfssone
  \newmathalphabet{\mathit}
  \addtoversion{normal}{\mathit}{cmr}{m}{it}
  \addtoversion{bold}{\mathit}{cmr}{bx}{it}
  \newmathalphabet{\mathbfit} 
  \addtoversion{normal}{\mathbfit}{cmr}{bx}{it}
  \addtoversion{bold}{\mathbfit}{cmr}{bx}{it}
  \newmathalphabet{\mathbfss} 
  \addtoversion{normal}{\mathbfss}{cmss}{bx}{n}
  \addtoversion{bold}{\mathbfss}{cmss}{bx}{n}
  \ifAMStwofonts
    \ifCUPmtlplainloaded \else
      \UseAMStwoboldmath
      \makeatletter
      \new@mathgroup\upmath@group
      \define@mathgroup\mv@normal\upmath@group{eur}{m}{n}
      \define@mathgroup\mv@bold\upmath@group{eur}{b}{n}
      \edef\UPM{\hexnumber\upmath@group}
      \new@mathgroup\amsa@group
      \define@mathgroup\mv@normal\amsa@group{msa}{m}{n}
      \define@mathgroup\mv@bold\amsa@group{msa}{m}{n}
      \edef\AMSa{\hexnumber\amsa@group}
      \makeatother
      \mathchardef\upi="0\UPM19
      \mathchardef\umu="0\UPM16
      \mathchardef\upartial="0\UPM40
      \mathchardef\leqslant="3\AMSa36
      \mathchardef\geqslant="3\AMSa3E
    \fi
  \fi
\fi 

\ifnfsstwo
  \DeclareMathAlphabet{\mathbfit}{OT1}{cmr}{bx}{it}
  \SetMathAlphabet\mathbfit{bold}{OT1}{cmr}{bx}{it}
  \DeclareMathAlphabet{\mathbfss}{OT1}{cmss}{bx}{n}
  \SetMathAlphabet\mathbfss{bold}{OT1}{cmss}{bx}{n}
  \ifAMStwofonts
    \ifCUPmtlplainloaded \else
      \DeclareSymbolFont{UPM}{U}{eur}{m}{n}
      \SetSymbolFont{UPM}{bold}{U}{eur}{b}{n}
      \DeclareSymbolFont{AMSa}{U}{msa}{m}{n}
      \DeclareMathSymbol{\upi}{0}{UPM}{"19}
      \DeclareMathSymbol{\umu}{0}{UPM}{"16}
      \DeclareMathSymbol{\upartial}{0}{UPM}{"40}
      \DeclareMathSymbol{\leqslant}{3}{AMSa}{"36}
      \DeclareMathSymbol{\geqslant}{3}{AMSa}{"3E}
    \fi
  \fi
\fi 

\ifCUPmtlplainloaded \else
  \ifAMStwofonts \else 
    \def\upi{\pi}
    \def\umu{\mu}
    \def\upartial{\partial}
  \fi
\fi

\title{The performance of spherical wavelets to detect non-Gaussianity in the 
CMB sky}
\author[E. Mart\'\i nez-Gonz\'alez et al.] {E. Mart\'\i nez-Gonz\'alez$^{1}$, J.E. Gallegos$^{1}$,
F. Arg\"ueso$^{2}$, L. Cay\'on$^{1}$, J.L. Sanz$^{1}$\\ 
$^1$Instituto de F\'\i sica de Cantabria, Fac. Ciencias, Av. los
	Castros s/n, 39005 Santander, Spain\\
$^2$Dpto. de Matem\'aticas, Universidad de Oviedo, c/ Calvo Sotelo s/n, 33007 Oviedo, Spain}
\date{\today}
\pagerange{\pageref{firstpage}--\pageref{lastpage}}
\pubyear{2001}

\begin{document}

\label{firstpage}

\maketitle


\begin{abstract}
We investigate the performance of spherical wavelets in discriminating 
between standard inflationary models (Gaussian) and non-Gaussian models. For
the later we consider small perturbations of the Gaussian model in
which an artificially specified skewness or kurtosis is introduced
through the Edgeworth expansion. By combining all the information
present in all the wavelet scales with the Fisher discriminant, we
find that the spherical Mexican Hat wavelets are clearly superior to
the spherical Haar wavelets. The former can detect levels of the
skewness and kurtosis of $\approx 1\%$ for $33'$ resolution, an order of 
magnitude smaller than the
later. Also, as expected, both wavelets are
better for discriminating between the models than the direct
consideration of moments of the temperature maps. The introduction of
instrumental white noise in the maps, $S/N=1$, does not change the
main results of this paper.  
\end{abstract}


\section{Introduction}

Most of the analyses of Cosmic Microwave Background (CMB) data focus on the 
measurement of the power spectrum of temperature
fluctuations. Information on this second order moment is crucial to
determine the fundamental parameters of the cosmological model
corresponding to our universe. However, this determination relies on the 
Gaussian hypothesis for the temperature distribution. Establishing the 
statistical
character of the CMB fluctuations will provide crucial evidence about
the physical origin of the primordial density fluctuations in the
early universe. Simple inflationary models predict a Gaussian,
homogeneous and isotropic random field for the temperature
fluctuations. On the contrary, non-standard inflation and cosmic
defects generically predict non-Gaussian random fields. Recent
CMB observations by Boomerang, DASI and MAXIMA-1 (Netterfield et
al. 2001, Pryke et al. 2001, Stompor et al. 2001) have established for 
the first time the 
presence of multiple acoustic peaks in the CMB power spectrum. As a
consequence cosmic defects cannot be the dominant source of density
perturbations in the universe. Even if they are present as a
sub-dominant component confirmation of its existence will be best made 
by appropriate techniques searching for non-Gaussian features in the
CMB maps.

Since a random field can departure from a Gaussian one in many different ways 
there is not a unique way to detect and characterise deviations from 
Gaussianity. Thus, depending on the kind of features one is looking
for some specific methods will prove to be more efficient than others.
Efficient methods are able
to extract relevant information on the non-Gaussian nature of the data 
which is otherwise hidden in the temperature fluctuation maps. A large
number of methods have been already proposed to search for
non-Gaussianity in CMB maps. The methods can be grouped by the spaces
(real, Fourier,...) in which they act. In real 
space, standard quantities used are the cumulants which contain information 
on the 1-pdf only. Information on the n-pdf can be obtained through
the Edgeworth expansion (Contaldi et al. 2000) or alternative
expansions with a proper normalization (Rocha et al. 2000). Other
quantities focus on topological and geometric statistics,
e.g. Minkowski functionals implemented on the sphere (Schmalzing and
Gorski 1998);
statistics of excursion sets, e.g. characteristics of peaks (Barreiro, 
Mart\'\i nez-Gonz\'alez and Sanz 2001), extrema correlation function
(Heavens and Gupta 2001); also geometrical characteristics of
polarisation have already been investigated (Naselsky and Novikov
1998). Multifractal analysis and roughness have
been applied to the COBE-DMR data (Diego et al. 1998, Mollerach et
al. 1999). In Fourier space, the bispectrum has been applied in
several occasions to analyse the COBE-DMR data (see e.g. Ferreira et
al. 1998) as well as an extension to include possible correlations among
multipoles (Magueijo 2000). An alternative approach is to work in eigen space,
extracting the eigenmodes from a principal component analysis. This
approach has been taken by Bromley and Tegmark (1999) for the COBE-DMR 
data and by Wu et al. (2001) for the MAXIMA-1 data. In spite of all
this effort there is not any
strong evidence of deviations from Gaussianity in the CMB up to date
(see however Magueijo 2000 for a possible deviation). More definitive 
conclusions about the statistical distribution of the CMB fluctuations 
are expected from data analyses of present and future sensitive 
experiments at arcmin resolution.

In this work we concentrate on wavelet analyses.
As it is often pointed out, wavelets are a very useful tool for data
analysis due to its space-frequency localisation. It has been already
demonstrated in many applications in a wide variety of scientific
fields. In particular in relation to the CMB the COBE-DMR data has
been studied with several
wavelet bases acting on the faces of the quad-cube COBE
pixelisation (Pando et al 1998, Mukherjee et al. 2000, Aghanim et
al. 2001). More appropriate analyses should involve the use of
spherical wavelets as in Tenorio et al. (1999). More recently Barreiro 
et al. (2000) and Cay\'on et al. (2001) have convolved the COBE-DMR
data with spherical wavelets in the HEALPiX pixelisation (Gorski,
Hivon \& Wandelt 1999) to test the Gaussianity of these data.
Those works
used the Spherical Haar Wavelet (SHW) and the Spherical Mexican Hat
Wavelet (SMHW), respectively. 

It is our aim in this work to confront the performance 
of these two spherical wavelet bases proposed  for discriminating
between standard inflationary (Gaussian) models and non-Gaussian models which
contain artificially specified moments (skewness or kurtosis) in the
temperature distribution. Physically
motivated non-Gaussian features can enter in the CMB maps in many
ways. Cosmic defects can produce linear discontinuities (cosmic
strings, Kaiser and Stebbins 1986), hot spots (global monopoles,
Coulson et al. 1994) or cold and hot spots (cosmic
textures, Turok and Spergel 1992). Non-standard inflationary models,
e.g. with several interacting scalar fields, are
expected to produce a qualitatively different non-Gaussianity. In
particular, models with an extra quadratic term in
the potential (Linde and Mukhanov 1997) generate a clear signal in the 
third moment (Verde et al. 2000, Komatsu and Spergel 2001). In any
case, it is very difficult to imagine a non-Gaussian primordial model 
producing no significant amount of neither of the two low order moments.

The paper is structured as follows. In Section 2 we introduce the Spherical 
Mexican Hat Wavelets (SMHW). Section 3 summarizes the main properties of the 
Spherical Haar Wavelets (SHW) and the procedure to calculate their 
coefficients. All-sky simulated non-Gaussian CMB maps at arcmin resolution, 
with a given power spectrum and artificially specified skewness or kurtosis, 
are generated in section 4. In section 5 we present optimal
statistics based on the wavelet coefficients to get a maximum
discriminating power between the Gaussian and non-Gaussian temperature 
maps. The main results are given in section 6 and we summarize the main 
conclusions of the paper in section 7.

\section{The spherical Mexican Hat wavelets}

Future CMB missions will provide temperature data covering all or almost all 
the sphere at arcmin resolution. It is thus necessary to have convenient 
pixelisation of the sphere which allows efficient analyses of the data. 
Wavelets defined on the plane have been widely used in astrophysical 
applications during the last years. In particular, the Mexican Hat
wavelet family has been successfully used to extract point 
sources from CMB maps (Cay\'on et al. 2000, Vielva et
al. 2001). However, applications of spherical 
wavelets have been very scarce and limitted to a few families of wavelets. 
Below we describe a procedure to extend the Mexican Hat wavelets to the
sphere.      

\subsection{The MEXHAT on $R^2$}

  A continuous ${\it wavelet}$ family on the plane $R^2$ is a set of filters
built from a mother wavelet $\psi(\vec{x})$, 
$\Psi (\vec{x}; \vec{b}, R) = \frac{1}{R}\psi (\frac{|\vec{x} - \vec{b}|}{R})$ 
(we only consider isotropic wavelets). $\psi(x)$ satisfies the 
following conditions:

\begin{equation}
\int d\vec{x}\,\psi (x) = 0\ \ \  (compensation),\ \ \ 
\end{equation}

\begin{equation}
C_{\psi}\equiv {(2\pi )}^2\int dq\,q^{-1}{\psi}^2(q) < \infty\ \ \ (admissibility),\ \ \ 
\end{equation}

\noindent where $\vec{b}$ defines a translation and $R$ a scale, i. e. we consider a
$3$-parameter family of filters. $\psi (q)$ is the 
Fourier transform of $\psi$ and we have introduced the standard normalization 

\begin{equation}
\int d\vec{x}\,{\Psi}^2 (\vec{x}; R) = \frac{1}{R^2}\int d\vec{x}\,{\psi}^2 (x) = 1,\ \ \ 
x\equiv |\vec{x}|.
\end{equation}
\vskip 0.3cm

\noindent {\bf a) Analysis}
 
Let us consider a function on the plane $f(\vec{x})$. The continuous
wavelet transform with respect to $\Psi$ is defined as the linear operation

\begin{equation}
w(\vec{b}, R) = \int d\vec{x}\,f(\vec{x})\Psi (\vec{x}; \vec{b}, R)
= \frac{1}{R}\int d\vec{x}\,f(\vec{x} + \vec{b})\psi (x/R).\ \ \ 
\end{equation}

\noindent $w(\vec{b}, R)$ are the wavelet coefficients dependent on $3$ parameters.
\vskip 0.3cm

\noindent {\bf b) Synthesis}  
 
 It can be proven that for any $\psi$ the following equality holds
 
\begin{equation}
\int dR\,d\vec{b}\,R^{-5} \psi (\frac{|\vec{x} - \vec{b}|}{R})
\psi (\frac{|\vec{x^{\prime}} - \vec{b}|}{R}) = C_{\psi}\delta (\vec{x} - \vec{x^{\prime}}),
\end{equation}

\noindent where $\delta (\vec{x})$ is the Dirac distribution.
 
 A straightforward calculation based on the previous equation leads to the 
continuous
reconstruction formula
 
\begin{equation} 
f(\vec{x}) = \frac{1}{C_{\psi}}\int dR\,d\vec{b}\,R^{-4}w(\vec{b}, R) 
\psi (\frac{|\vec{x} - \vec{b}|}{R}).
\end{equation} 
\vskip 0.3cm

\noindent {\bf c) The MEXHAT wavelets}
 
 A particular example is the MEXHAT wavelet defined by

\begin{equation}
\Psi (x; R)\equiv \Psi (\vec{x}; \vec{0}, R) = \frac{1}{{(2\pi )}^{1/2}R}
[2 - {(\frac{x}{R})}^2]e^{- x^2/2R^2},\\
x\equiv |\vec{x}|.
\end{equation}

\noindent This wavelet (introduced by Marr (1980)) is proportional to the 2D Laplacian of the
Gaussian function. It has been extensively used in the literature to detect structure on a 2D 
image (e.g. in astrophysics to detect point sources in a noisy background).

\subsection{The MEXHAT on $S^2$}

For CMB analyses we are interested in the extension of these isotropic 
wavelets to the sphere.
Recently, Antoine \& Vandergheynst (1998) have followed a group theory approach to deal 
with this problem. This extension incorporates four basic properties: a) the basic function is a compensated
filter, b) translations, c) dilations and d) Euclidean limit for small angles. They
conclude that the stereographic projection on the sphere is the appropriate one to translate the 
mentioned properties from the plane to the sphere. Such a projection is defined by 
$(\vec{x})\mapsto (\theta ,\phi )$

\begin{equation}
x_1 = 2\tan \frac{\theta}{2}\cos \phi ,\ \ \ 
x_2 = 2\tan \frac{\theta}{2}\sin \phi ,
\end{equation}

\noindent where $(\theta ,\phi )$ represent polar coordinates on $S^2$ and 
$(y\equiv 2\tan \frac{\theta}{2}, \phi)$ are polar coordinates in the tangent plane to the North
pole (see Figure 1).

\begin{figure}
\epsfxsize=8cm
\begin{minipage}{\epsfxsize}\epsffile{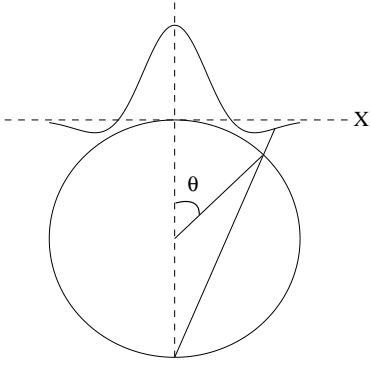}\end{minipage}
\caption{ Stereographic projection to translate the
properties of the Mexican Hat Wavelet from the plane to the sphere}
\label{Stereographi}
\end{figure}

Therefore, the isotropic wavelet $\Psi (x; R)$ transforms to

\begin{equation}
\Psi_{S} (\theta; R) \propto {(\cos \frac{\theta}{2})}^{-4}\Psi (x\equiv 2\tan \frac{\theta}{2}; R).
\end{equation}

It can be proven that the new wavelet on $S^2$ incorporates the basic properties, i. e.
a) it is a compensated filter ($\int d\theta d\phi \sin \theta \Psi_{S} (\theta; R) = 0$), 
b) translations are
defined by translations along the sphere, i. e. rotations about the
center of the sphere, c) the
dilations are defined by the stereographic projection of dilations on
the plane and d) for small angles one recovers the Euclidean limit.
\vskip 0.3cm

\noindent {\bf a) Analysis}
 
  Let us consider a function on the sphere $f(\theta, \phi)$. The continuous wavelet transform
with respect to $\Psi_S (\theta ; R)$ is defined as the linear operation

\begin{equation}
\tilde{w}(\vec{x}, R) = \int d\theta^{\prime}\,d\phi^{\prime}\sin \theta^{\prime}
\,\tilde{f}(\vec{x} + \vec{\mu})\Psi_S (\theta^{\prime}; R).\ \ \ 
\end{equation}
\begin{eqnarray}
\vec{x} \equiv 2\tan \frac{\theta}{2}(\cos \phi , \sin \phi ),\ \ \ 
\vec{\mu} \equiv 2\tan \frac{\theta^{\prime}}{2}(\cos \phi^{\prime} , \sin \phi^{\prime}),{} \nonumber\\
\tilde{f}(\vec{x})\equiv f(\theta ,\phi ){},
\end{eqnarray}

\noindent $w(\theta ,\phi ; R)\equiv \tilde{w}(\vec{x}, R)$ are the wavelet coefficients 
dependent on $3$ parameters.
\vskip 0.3cm

\noindent {\bf b) Synthesis}  
  
 A straightforward calculation based on the equation (5) leads, after stereographic projection, 
to the continuous reconstruction formula:
 
\begin{eqnarray} 
f(\theta ,\phi )\equiv\tilde{f}(\vec{x}) = \frac{1}{C_{\psi}}
\int d\theta^{\prime}d\phi^{\prime}\sin\theta^{\prime}\frac{dR}{R^3}
\tilde{w}(\vec{x} + \vec{\mu}, R){} 
\nonumber\\ 
\Psi_S (\theta^{\prime}; R){}\ ,
\end{eqnarray} 

\noindent where $\tilde{w}(\vec{x}, R)\equiv w(\theta ,\phi ; R)$.
\vskip 0.3cm

\noindent {\bf c) The MEXHAT wavelets}
 
\begin{figure}
\epsfxsize=8cm
\begin{minipage}{\epsfxsize}\epsffile{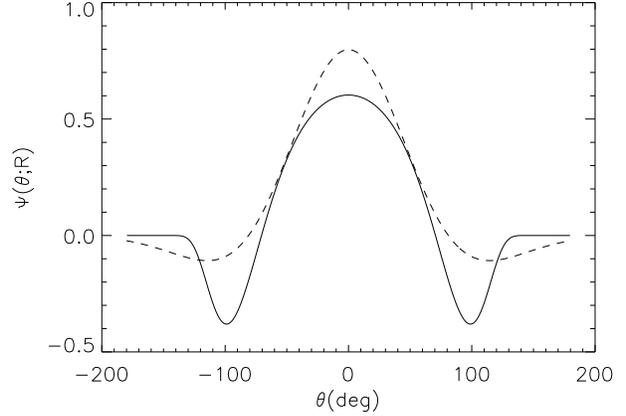}\end{minipage}
\caption{Mexican Hat Wavelet as deformed on the sphere (solid line)
from the plane (dashed line). The scale of the wavelet is chosen to be
$R=1$ rad.} 
\label{MexHatSphere}
\end{figure}

A particular example is the MEXHAT wavelet defined by (see Figure 
\ref{MexHatSphere})

\begin{equation}
\Psi (\theta; R) = \frac{1}{{(2\pi)}^{1/2}R\,N}{[1+{(\frac{y}{2})}^2]}^2
[2 - {(\frac{y}{R})}^2]e^{-{y}^2/2R^2},\ \ \ 
\end{equation}

\begin{equation}
N(R)\equiv {(1 + \frac{R^2}{2} + \frac{R^4}{4})}^{1/2},
y\equiv 2\tan \frac{\theta}{2}.
\end{equation}

We remark that the normalization constant has been chosen such that 
$\int d\theta d\phi \sin \theta \Psi^2 (\theta; R) = 1$. This is the
wavelet we are going to use
in this paper to analize non-Gaussianity associated to different models.

We comment that the stereographic projection of the MEXHAT wavelet has
been recently used to
analize maps of the cosmic microwave background radiation (CMB)
(Cay\'on et al. 2001).

\section{Spherical Haar Wavelets}

SHW were introduced by Sweldens (1995) as a generalization of planar Haar
wavelets to the pixelised sphere. They are orthogonal and adapted to a
given pixelisation of the sky which must be hierarchical, contrary to
the SMHW which are non-orthogonal and redundant. However they
are not obtained from
dilations and translations of a mother wavelet, contrary to planar
Haar wavelets and SMHW. As for the planar Haar wavelets, they
possess a good space-frequency localisation. However, their frequency
localisation is not as good as that of the SMHW. 
Two applications of SHW to the analysis of CMB maps have already been
performed. Tenorio et al. (1999) apply them to simulated CMB skies on
the QuadCube pixelisation. They study the CMB spatial structure by
defining a position-dependent measure of power. Also they show their
efficiency in denoising and compressing CMB data. Barreiro et
al. (2000) tested the Gaussianity of the COBE-DMR data on the HEALPix
pixelisation. One of the advantages of HEALPix over
QuadCube is that there is no need to correct for the pixel area. 

Since detailed description of the SHW transform has already been given
in the previous papers, we here describe the main features of the
wavelet decomposition. The SHW decomposition is based on one scaling
$\phi_{j,k}$ and three wavelet functions $\psi_{m,j,k}$ at each
resolution level j and position on the grid k. For HEALPix the resolution
is given in terms of the number of divisions in which each side of the
basic 12 pixels is divided, $N_{side}=2^{j-1}$. Thus, for level $j$
the total number
of pixels with area $\mu_j$ is given by $n_j=12\times 4^{j-1}$. Each
pixel k at resolution j, $S_{j,k}$ is divided into four pixels
$S_{j+1,k_0}, ..., S_{j+1,k_3}$ at resolution j+1. For
computational reasons the maximum resolution we will consider in our
simulations is $J=9$ which corresponds to $N_{side}=256$. The scaling
and wavelet functions are simply given by

\begin{equation}
\phi_{j,k}(x)=\left\{\begin{array}{ll}
1 & if x\in S_{j,k}\\
0 & otherwise\ ,
\end{array} \right.
\end{equation}
\begin{equation}
\psi_{1,j,k}=\frac {\phi_{j+1,k_0}+\phi_{j+1,k_2}-\phi_{j+1,k_1}-
\phi_{j+1,k_3}}{4\mu_{j+1}}
\end{equation}
\begin{equation}
\psi_{2,j,k}=\frac {\phi_{j+1,k_0}+\phi_{j+1,k_1}-\phi_{j+1,k_2}-
\phi_{j+1,k_3}}{4\mu_{j+1}}
\end{equation}
\begin{equation}
\psi_{3,j,k}=\frac {\phi_{j+1,k_0}+\phi_{j+1,k_3}-\phi_{j+1,k_1}-
\phi_{j+1,k_2}}{4\mu_{j+1}}
\end{equation}
where $k_0,k_1,k_2,k_3$ are the four pixels at resolution level $j+1$
in which the pixel $k$ at level $j$ is divided. Please note that the three
wavelet functions so defined differ from the ones used by Tenorio et
al (1999) and Barreiro et
al. (2000). We choose those expressions by similarity with the diagonal,
vertical and horizontal details defined on the plane. The
reconstruction of the temperature field is obtained by
\begin{eqnarray}
\frac {\Delta T}{T} (x_i)=\sum_{l=0}^{n_{j_0}-1}\lambda_{j_0, l}\phi_{j_0,j}
(x_i) + \sum_{m}\sum_{j=j_0}^{J-1}\sum_{l=0}^{n_j-1}\gamma_{m,j,l} 
\nonumber\\
\psi_{m,j,l}(x_i)\ ,
\end{eqnarray}
where $\lambda_{j_0,k}$ and $\gamma_{m,j,k}$ are the approximation and
detail coefficients respectively. The level index $j$ goes from the
finest resolution $J$ to the coarsest one considered $j_0$.  

\begin{figure}
\epsfxsize=8cm
\begin{minipage}{\epsfxsize}\epsffile{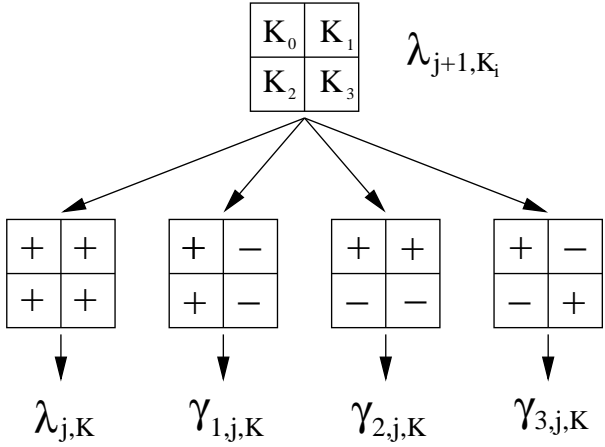}\end{minipage}
\caption{Hierarchical structure of Wavelet coefficients for the
Spherical Haar Wavelet.} 
\label{HaarWave}
\end{figure}

The wavelet coefficients at level $j$ can be obtained from the four
corresponding approximation coefficients at level $j+1$, $\lambda_{j+1,k_i}$ as 
follows (see figure \ref{HaarWave}.):
\begin{equation}
\lambda_{j,k}=\frac {1}{4}\sum_{i=0}^{3} \lambda_{j+1,k_i}
\end{equation}
\begin{equation}
\gamma_{1,j,k}=\mu_{j+1}(\lambda_{j+1,k_0}+\lambda_{j+1,k_2}-
\lambda_{j+1,k_1}-\lambda_{j+1,k_3})
\end{equation}
\begin{equation}
\gamma_{2,j,k}=\mu_{j+1}(\lambda_{j+1,k_0}+\lambda_{j+1,k_1}-
\lambda_{j+1,k_2}-\lambda_{j+1,k_3})
\end{equation}
\begin{equation}
\gamma_{3,j,k}=\mu_{j+1}(\lambda_{j+1,k_0}+\lambda_{j+1,k_3}-
\lambda_{j+1,k_1}-\lambda_{j+1,k_2})\ \ .   
\end{equation}
The generation of coefficients start with the original map, finest
resolution $j=J$, for which the coefficients $\lambda_{J,k}$ are
identified with the temperature fluctuation at  pixel $k$.

Finally, from the definition of the SHW it is easily seen that this
wavelet is not rotationally invariant, contrary to the SMHW.

\section{Non-Gaussian simulations}

There are many ways in which
physically motivated non-Gaussian features can enter in the CMB
temperature distribution. However, up to now there is no evidence of their
existence, being all experimental data consistent with Gaussianity
(Kogut et al. 1996, Barreiro et al. 2000, Aghanim et al. 2000, Cay\'on
et al. 2001, Wu et al. 2001; see however Magueijo 2000 for a possible
positive signal in the COBE-DMR data, although that detection has not
been confirmed by any of the other analyses). If 
departures from Gaussianity of cosmological origin really exist they
will more likely be small and all-sky, sensitive, arcminute resolution
experiments will be needed for their detection.

Here the spherical wavelets will be tested against 
non-Gaussian simulations of artificially
specified moments that will be assumed to be small. In this
case a useful way to construct non-Gaussian distributions is by
perturbing the Gaussian one through a sum of moments, the
Edgeworth expansion. 
For simplicity we will consider the two lowest cumulants to
characterise the deviations from normality: skewness and kurtosis.
As discussed in the introduction alternative models to standard
inflation, e.g. cosmic defects as a subdominant source of density
perturbations or non-standard inflation, can produce 
significant levels of at least one of the two moments.

\subsection{Edgeworth expansion}

For small deviations from Gaussianity, there is a wide class of
distributions that can be given in terms of a Gaussian distribution
times an infinite sum of its cumulants. This is the well known
Edgeworth expansion. The problem with this expansion is that setting
all cumulants to zero except one does not guarantee the 
positive definiteness and normalization that a distribution has to satisfy.   
However, for small
deviations from normality the resulting function is always positive at
least up to many sigmas in the tail of the distribution and 
the normalization factor required for the function to become a well
defined distribution is very small and
does not appreciably disturb the non-zero moments (i.e. skewness or
kurtosis) introduced in the first place.

The Edgeworth expansion can be obtained from the characteristic function
$\phi(t)$ by considering the linear terms in the cumulants and performing the
inverse Fourier transform to recover the density function $f(x)$: 
\begin{equation}
f(x)=G(x)\bigl\{1+\sum_{n=3}^{\infty}\frac{k_n}{n!2^{n/2}}
H_n\bigl(\frac{x}{\sqrt{2}}\bigr)+O(k_nk_{n'})\bigr\}\ \ ,
\end{equation} 
where $H_n$ is the Hermite polynomial. 
Considering the perturbations corresponding to the skewness and
kurtosis and keeping only the first terms in the corresponding
Hermite polynomials, we have
\begin{equation}
f_S(x)=\frac{e^{-\frac{x^2}{2}}}{\sqrt{2\pi}}\Bigl(1+
\frac{S}{6}\bigl(x(x^2-3)\bigr)\Bigr)\ , 
\end{equation} 
\begin{equation}
f_K(x)=\frac{e^{-\frac{x^2}{2}}}{\sqrt{2\pi}}\Bigl(1+
\frac{K}{24}(x^4-6x^2+3)\Bigr)\ ,  
\end{equation}    
where S, K denote skewness and kurtosis, respectively. We will use
these equations to generate our artificially specified non-Gaussian 
distributions. Since the resulting
distribution is not well defined even for the case of small skewness
and kurtosis we set the function to zero when it becomes negative and
we also normalize it appropriately. We remark that the zero cuts of
the distribution, if present, appear far away in the tails of the 
distribution for the case of small values of skewness and kurtosis
that we consider here. Also, as a consequence, the normalization value
required is very close to 1. In this way we checked that the initial
values of the skewness and kurtosis we start with in the Edgeworth
expansion does not appreciably change after the necessary changes
introduced to obtain a well defined probability density function (pdf).

In order to make the simulations resemble the CMB data observed by
a given experiment we smooth them with a Gaussian filter. For
practical reasons we use a FWHM of $33'$ which  may correspond to some
of the channels in all-sky experiments like MAP and Planck 
(e.g. the $30$GHz channel of the Planck mission). We choose to work on
the HEALPix pixelisation of the sphere with a resolution
$N_{side}=256$. We use the HEALPix
package to perform the analysis of our simulated CMB data. However, it is
not adequate to use that package to convolve our unfiltered 
independent temperature data with the Gaussian $33'$ FWHM beam in Fourier 
space, instead we perform the convolution in real space. After that, in order
to make the simulations more realistic we
normalize the CMB power spectrum $C_l$ of both Gaussian and
non-Gaussian simulations to that of a CDM flat
$\Lambda$-model  using the HEALPix package. As a consequence of the 
beam convolution and the introduction of correlations in the temperature maps 
the original levels of 
skewness and kurtosis injected through the Edgeworth expansion are
reduced (compare columns 1 and 2 in table \ref{Fisher}).
The performance of
spherical wavelets will be tested with these simulations in section 5.

\subsection{Distribution of spherical wavelet coefficients}

Since wavelet coefficients represent linear transformations of the
original data, in the case of a Gaussian distribution the wavelet
coefficients remain Gaussian distributed. This a very nice property of
wavelets and all we have to do to test Gaussianity in wavelet space is
to look from deviations from normality. 

However, for the case of the sphere any given pixelisation scheme will
introduce biases. The specific bias introduced will depend on, for instance, 
whether the 
pixels are not of equal area or the distances between one pixel and its 
neighbours vary with the position on the
sphere. This is in fact the situation for the two pixelisations
already used to analyse all-sky CMB temperature fluctuations. For the
COBE-DMR experiment the pixelisation used was the Quad-Cube and in
this projection of the cube on the sphere equal-area pixels on the
sides of the cube appear with different area when projected on the
sphere. For present satellite experiments like MAP and Planck the HEALPix 
pixelisation is now widely used. While this pixelisation possesses very
nice properties, such as equal area iso-latitude pixels, however the
distances between one pixel and its neighbours vary with
latitude. Pixels near the equator
tend to be more uniformly distributed than those near the poles. As we will
compute in next section, this property produces a bias in the kurtosis
of the wavelet coefficients for the case of the SHW (see table 1, Gaussian 
case which corresponds to a null injected value for the kurtosis).
For the Gaussian and non-Gaussian simulations which will be performed
in next section we will compute the
first cumulants of the coefficients of the two spherical wavelets
considered in this paper for the HEALPix scheme. For the SHW the
coefficients correspond to
three different details: diagonal, vertical and horizontal. Since
those details are directly obtained from linear operations of the four
neighbour pixels (as we saw in the previous section) and pixels are not
equally separated all over the sphere,  
correlations present in the temperature fluctuations make the wavelet
coefficients to be biased.
This bias produces a peaked distribution with respect to a Gaussian and 
therefore a positive kurtosis in the three details of the SHW coefficients
even for temperature realizations derived from normal distributions
(as can be seen from table 1, the mean value of the
kurtosis for the finest resolution of the Gaussian model is displaced about 
$10\sigma$ from zero). 

In the case of the SMHW we only have a type of
coefficients for each scale. Since this is a continuous, rotationally
invariant wavelet -and thus not adapted to the pixelisation- no bias
is produced in this case. 
\begin{center}
\begin{table*}
\caption{Mean and standard deviation, within parenthesis, for the different 
wavelet scales.}
\label{meanstddev}
\scriptsize{
\begin{tabular}{cccccccccccccccccccccccccc}
\noalign{\smallskip}
\hline
Injected & Wavelet  &&SMHW&&&&&&&      &&& SHW  &&&&      &&&&&& Temperature& \\
         & Scale&&&&&      & vert &&&&&& diag &&&&&& hori &&&&&&\\
\hline
\end{tabular}
\begin{tabular}{c}
SKEWNESS\\
\end{tabular}
\begin{tabular}{ccrrrrr}  
\hline
 0.00 & 1 pix&-1.0$\times10^{-4}$(5.6$\times10^{-3}$)&-1.3$\times10^{-4}$(7.6$\times10^{-3}$)& 1.7$\times10^{-4}$(5.5$\times10^{-3}$)&-2.1$\times10^{-4}$(7.3$\times10^{-3}$)&-1.1$\times10^{-3}$(2.3$\times10^{-2}$)\\
      & 2 pix&-1.0$\times10^{-4}$(6.3$\times10^{-3}$)&-3.0$\times10^{-4}$(1.1$\times10^{-2}$)& 7.3$\times10^{-5}$(9.8$\times10^{-3}$)& 7.2$\times10^{-5}$(1.1$\times10^{-2}$)&\\
\hline
 0.00*& 1 pix&-2.1$\times10^{-5}$(3.4$\times10^{-3}$)& 2.1$\times10^{-4}$(5.6$\times10^{-3}$)&-3.1$\times10^{-4}$(5.3$\times10^{-3}$)&-1.0$\times10^{-5}$(5.5$\times10^{-3}$)&-4.8$\times10^{-4}$(6.7$\times10^{-3}$)\\
      & 2 pix&-1.7$\times10^{-4}$(5.9$\times10^{-3}$)&-7.8$\times10^{-4}$(1.1$\times10^{-2}$)& 8.8$\times10^{-4}$(1.1$\times10^{-2}$)& 1.1$\times10^{-4}$(1.1$\times10^{-2}$)&\\
\hline
 0.05 & 1 pix& 1.3$\times10^{-2}$(5.0$\times10^{-3}$)&-2.7$\times10^{-4}$(6.9$\times10^{-3}$)&-1.7$\times10^{-3}$(5.5$\times10^{-3}$)& 2.8$\times10^{-4}$(7.2$\times10^{-3}$)&9.0$\times10^{-3}$(2.4$\times10^{-2}$)\\
      & 2 pix& 7.5$\times10^{-3}$(6.1$\times10^{-3}$)&-1.6$\times10^{-3}$(1.1$\times10^{-2}$)&-4.7$\times10^{-4}$(9.3$\times10^{-3}$)& 2.0$\times10^{-6}$(1.2$\times10^{-2}$)& \\
\hline
 0.10 & 1 pix& 2.7$\times10^{-2}$(5.2$\times10^{-3}$)& 2.8$\times10^{-4}$(7.1$\times10^{-3}$)&-3.9$\times10^{-3}$(5.6$\times10^{-3}$)& 6.0$\times10^{-5}$(7.3$\times10^{-3}$)& 1.6$\times10^{-2}$(2.3$\times10^{-2}$)\\
      & 2 pix& 1.5$\times10^{-2}$(6.1$\times10^{-3}$)& 5.7$\times10^{-4}$(1.1$\times10^{-2}$)&-1.4$\times10^{-3}$(9.6$\times10^{-3}$)&-6.3$\times10^{-4}$(1.1$\times10^{-2}$)& \\
\hline
 0.30 & 1 pix &7.6$\times10^{-2}$(5.4$\times10^{-3}$)&2.6$\times10^{-4}$(7.3$\times10^{-3}$)&-1.0$\times10^{-2}$(5.6$\times10^{-3}$)& 3.8$\times10^{-4}$(7.7$\times10^{-3}$)& 4.6$\times10^{-2}$(2.4$\times10^{-2}$)\\
      & 2 pix &4.3$\times10^{-2}$(6.0$\times10^{-3}$)&6.7$\times10^{-5}$(1.1$\times10^{-2}$)&-3.3$\times10^{-3}$(9.5$\times10^{-3}$)& 5.6$\times10^{-4}$(1.1$\times10^{-2}$)& \\
\hline
 0.30*& 1 pix & 9.5$\times10^{-3}$(3.5$\times10^{-3}$)& 3.5$\times10^{-5}$(5.4$\times10^{-3}$)&-4.0$\times10^{-4}$(5.7$\times10^{-3}$)&-2.0$\times10^{-4}$(5.8$\times10^{-3}$)& 1.1$\times10^{-2}$(6.8$\times10^{-3}$)\\
      & 2 pix & 3.1$\times10^{-2}$(6.2$\times10^{-3}$)& 2.4$\times10^{-4}$(1.0$\times10^{-2}$)& 4.6$\times10^{-6}$(1.1$\times10^{-2}$)& 9.9$\times10^{-5}$(1.1$\times10^{-2}$)& \\
\hline
 0.50 & 1 pix & 1.2$\times10^{-1}$(5.4$\times10^{-3}$)&-5.6$\times10^{-4}$(7.6$\times10^{-3}$)&-1.6$\times10^{-2}$(5.6$\times10^{-3}$)&-6.4$\times10^{-6}$(7.4$\times10^{-3}$)& 6.9$\times10^{-2}$(2.4$\times10^{-2}$)\\
      & 2 pix & 6.6$\times10^{-2}$(6.2$\times10^{-3}$)&-5.9$\times10^{-4}$(1.1$\times10^{-2}$)&-5.4$\times10^{-3}$(9.6$\times10^{-3}$)& 2.1$\times10^{-5}$(1.1$\times10^{-2}$)&\\ 
\hline
\end{tabular}

\begin{tabular}{c}
KURTOSIS \\
\end{tabular}

\begin{tabular}{ccrrrrr}
\hline
0.00 &1 pix &-3.6$\times10^{-4}$(1.0$\times10^{-2}$)& 1.7$\times10^{-1}$(2.0$\times10^{-2}$)& 1.8$\times10^{-1}$(1.8$\times10^{-2}$)& 1.7$\times10^{-1}$(1.9$\times10^{-2}$)&-3.4$\times10^{-3}$(2.6$\times10^{-2}$)\\
     &2 pix &-4.1$\times10^{-4}$(1.2$\times10^{-2}$)& 1.0$\times10^{-1}$(2.7$\times10^{-2}$)& 3.9$\times10^{-2}$(2.5$\times10^{-2}$)& 1.0$\times10^{-1}$(2.7$\times10^{-2}$)&\\
\hline
 0.00*& 1 pix&-8.7$\times10^{-5}$(6.4$\times10^{-3}$)& 4.1$\times10^{-3}$(1.6$\times10^{-2}$)&-5.7$\times10^{-5}$(1.1$\times10^{-2}$)& 4.4$\times10^{-3}$(1.5$\times10^{-2}$)&-1.1$\times10^{-3}$(6.9$\times10^{-3}$)\\
      & 2 pix&-9.7$\times10^{-4}$(9.5$\times10^{-3}$)& 1.9$\times10^{-2}$(2.3$\times10^{-2}$)& 1.9$\times10^{-3}$(2.2$\times10^{-2}$)& 2.1$\times10^{-2}$(2.3$\times10^{-2}$)&\\
\hline
0.10 &1 pix & 9.9$\times10^{-3}$(1.0$\times10^{-2}$)& 1.7$\times10^{-1}$(1.9$\times10^{-2}$)& 1.8$\times10^{-1}$(1.8$\times10^{-2}$)& 1.7$\times10^{-1}$(2.0$\times10^{-2}$)& 3.2$\times10^{-3}$(2.6$\times10^{-2}$)\\
     &2 pix & 3.9$\times10^{-3}$(1.3$\times10^{-2}$)& 1.1$\times10^{-1}$(2.7$\times10^{-2}$)& 4.1$\times10^{-2}$(2.6$\times10^{-2}$)& 1.1$\times10^{-1}$(2.8$\times10^{-2}$)& \\
\hline
0.30 &1 pix & 2.9$\times10^{-2}$(1.0$\times10^{-2}$)& 1.8$\times10^{-1}$(2.0$\times10^{-2}$)& 1.9$\times10^{-1}$(1.8$\times10^{-2}$)& 1.8$\times10^{-1}$(1.9$\times10^{-2}$)& 7.7$\times10^{-3}$(2.7$\times10^{-2}$)\\
     &2 pix & 1.2$\times10^{-2}$(1.3$\times10^{-2}$)& 1.1$\times10^{-1}$(2.7$\times10^{-2}$)& 4.8$\times10^{-2}$(2.6$\times10^{-2}$)& 1.1$\times10^{-1}$(2.8$\times10^{-2}$)& \\
\hline
0.40 &1 pix & 3.8$\times10^{-2}$(1.1$\times10^{-2}$)& 1.9$\times10^{-1}$(2.0$\times10^{-2}$)& 1.9$\times10^{-1}$(1.8$\times10^{-2}$)& 1.9$\times10^{-1}$(2.0$\times10^{-2}$)& 1.1$\times10^{-2}$(2.7$\times10^{-2}$)\\
     &2 pix & 1.7$\times10^{-2}$(1.3$\times10^{-2}$)& 1.1$\times10^{-1}$(2.8$\times10^{-2}$)& 5.3$\times10^{-2}$(2.6$\times10^{-2}$)& 1.2$\times10^{-1}$(2.8$\times10^{-2}$)& \\
\hline
0.50 &1 pix & 4.8$\times10^{-2}$(1.1$\times10^{-2}$)& 1.9$\times10^{-1}$(2.0$\times10^{-2}$)& 2.0$\times10^{-1}$(1.8$\times10^{-2}$)& 1.9$\times10^{-1}$(2.0$\times10^{-2}$)& 1.4$\times10^{-2}$(2.6$\times10^{-2}$)\\
     &2 pix & 2.1$\times10^{-2}$(1.3$\times10^{-2}$)& 1.2$\times10^{-1}$(2.8$\times10^{-2}$)& 5.3$\times10^{-2}$(2.5$\times10^{-2}$)& 1.2$\times10^{-1}$(2.8$\times10^{-2}$)& \\
\hline
 0.50*& 1 pix & 2.8$\times10^{-3}$(6.2$\times10^{-3}$)& 1.6$\times10^{-3}$(1.1$\times10^{-2}$)&-5.7$\times10^{-4}$(1.2$\times10^{-2}$)&-8.1$\times10^{-4}$(1.1$\times10^{-2}$)& 2.3$\times10^{-3}$(7.2$\times10^{-3}$)\\
      & 2 pix & 1.2$\times10^{-2}$(9.1$\times10^{-3}$)& 2.3$\times10^{-2}$(2.2$\times10^{-2}$)& 1.4$\times10^{-3}$(2.2$\times10^{-2}$)& 2.3$\times10^{-2}$(2.4$\times10^{-2}$)& \\
\hline
\end{tabular}
\begin{tabular}{l}
$^*$These models include the addition of noise to the maps with S/N = 1.\\ 
\end{tabular}
}
\normalsize
\rm
\end{table*}
\end{center}
\section{Discriminating power}      

The discriminating power of the spherical wavelets will be tested 
using Gaussian and non-Gaussian simulations with different amounts of
either skewness or kurtosis introduced using the Edgeworth
expansion, and normalized to a power spectrum $C_l$ consistent with
observations (as discussed above). 
Since
the skewness and kurtosis are introduced at the highest resolution
through the Edgeworth expansion (as described above), we
expect to detect them with the skewness and kurtosis of the spherical
wavelet coefficients also at
the highest resolutions. Thus we will consider for the analysis the
first five  resolution scales starting from the finest one. The scales
go as powers of 2 for the SHW and for comparison we choose the same
values for the SMHW parameter $R$: 1, 2, 4, 8 and 16 pixels. We can relate the
scales of the two wavelets by looking to the scaling functions. The relation 
between the side, $s$, of the step function (scaling 
function for the Haar wavelet) and the dispersion $R$ of the Gaussian is: 
$s=\sqrt{2\pi} R$. Then, for the finest scale $s=2$ pixels, which corresponds 
to an $R\approx 0.8$ pixels which is approximately 1 pixel. 

Results obtained in Fourier
space are equivalent to those obtained in real space if the
functions considered are bandwidth limited (with the bandwidth included in
the one covered by the pixelisation). We have checked this for the finest 
resolution of the SMHW. The average difference between the SMHW coefficients
computed by direct convolution in 
real space and going to Fourier space is $< 1\%$.

Given the 5 values of skewness or kurtosis corresponding to the 5
resolution scales for the SMHW and the 15 values for the
SHW (5 scales for each of the 3 details), we would like to construct a test
statistic which, combining all this information, can best distinguish
between the two hypotheses: a) $H_0$: the
data are drawn from a Gaussian model, b) $H_1$: the data are drawn
from a non-Gaussian model with either skewness or kurtosis. The best
test statistic in the sense of maximum power for a given significance
level is given by the likelihood ratio:
\begin{equation}
t(\vec{x})=\frac{f(\vec{x}|H_0)}{f(\vec{x}|H_1)}
\end{equation}
where $f(\vec{x}|H_0)$ and $f(\vec{x}|H_1)$ are the pdf of the data
given hypotheses $H_0$ and $H_1$, respectively. Since we do not know
those multivariate pdf´s and would be tremendously costly in cpu time
to determine
them by Monte Carlo simulations, we use as test statistic the simpler 
Fisher linear discriminant function (Fisher 1936; see also Cowan 1998). This
discriminant has been recently used by Barreiro and Hobson (2001) to
study the discriminanting power of planar wavelets to detect
non-Gaussianity in the CMB in small patches of the sky. The Fisher
discriminant is a linear function of the data that maximizes the
distance between the two pdf's, $g(t|H_0)$ and $g(t|H_1)$, such a distance 
defined as
the ratio $(\tau_0-\tau_1)^2/(\Sigma_0^2+\Sigma_1^2)$. $\tau_k$ and 
$\Sigma_k^2$,
$k=0,1$, are the mean and the variance of $g(t|H_k)$, respectively. The Fisher
discriminant is given by:
\begin{equation} 
t(\vec{x})=(\vec\mu_0-\vec\mu_1)^TW^{-1}\vec{x}
\end{equation}
with $W=V_0+V_1$ and $V_k$ the covariance matrix and $\vec\mu_k$ the mean 
values of
$f(\vec{x}|H_k)$. In the particular case that $f(\vec{x}|H_0)$ and 
$f(\vec{x}|H_1)$  are both multidimensional Gaussians with the same covariance
matrix, the Fisher discriminant is equivalent to  the likelihood
ratio. 
 
The mean values and covariance matrices of the skewness and kurtosis
at each resolution level for the Gaussian and non-Gaussian models are 
obtained from a large number of simulations. In the next section we
use those simulations to compare the power of the test $p\equiv
1-\beta$ to discriminate against the alternative hypothesis $H_1$ at
a given significance level $\alpha$ for the two
spherical wavelets. $\alpha$ and $\beta$ account for the probability
of rejecting the null hypothesis $H_0$ when it is actually true (error
of the first kind) and the probability of accepting $H_0$ when the
true hypothesis is $H_1$ and not $H_0$ (error of the second kind),
respectively. The decision to accept or reject $H_0$ is done by
defining a critical region for the statistic $t$; if the value of $t$
is greater than a cut value $t_{cut}$ the hypothesis $H_0$ is
rejected. Thus, $\alpha$ and $\beta$ are given by:
\begin{equation} 
\alpha=\int_{t_{cut}}^\infty dt g(t|H_0) \ ,
\end{equation}
\begin{equation} 
\beta=\int_{-\infty}^{t_{cut}} dt g(t|H_1) \ .
\end{equation}
This kind of analysis is very much along the lines
of the one performed by Barreiro and Hobson (2001) for planar
wavelets. From now on a value for
the sensitivity of $\alpha=1\%$ will be adopted.

\section{Results}  
\begin{center}
\begin{table}
\caption{Power of the Fisher discriminant at 1\% significance level}
\label{Fisher}
\scriptsize{
\begin{tabular}{llrrrr}
\noalign{\smallskip}
\hline
&Injected & True$^1$ & SMHW & SHW & Temperature\\
&&$\times10^{-2}$ & P(\%)&P(\%)&P(\%) \\
\hline
&0.05& 0.9(2.4)  & 66.8 & 1.51 & 2.51 \\
&0.10& 1.6(2.3) & 100 & 7.09  & 4.67  \\
SKEWNESS&0.30& 4.6(2.4) & 100 & 36.12 & 36.85\\
&0.30$^2$& 1.1(0.7)& 99.6 & 1.80& 2.83 \\
&0.50& 6.9(2.4) & 100 & 78.46 & 73.6\\
\hline
&0.10&0.3(2.6)&15.35&3.00&1.42\\
&0.30&0.8(2.7)&86.89&9.00&3.40\\
KURTOSIS&0.40&1.1(2.7)&98.10&16.11&4.90\\
&0.50&1.4(2.6)&99.90&28.43&3.50\\
&0.50$^2$&0.2(0.7)&20.84&1.00&0.32\\
\hline
\end{tabular}
\begin{list}{}{}
\item[$^1$]{True refers to the mean value obtained in the analysed maps. The 
standard deviation is given within parenthesis.}
\item[$^2$]{These models include the addition of noise to the maps with S/N = 1.} 
\end{list}
}
\normalsize
\rm
\end{table}
\end{center}
For both, Gaussian and not Gaussian models, we perform a $1000$
simulations. As commented above, to make the simulations more
realistic each simulation is convolved with a Gaussian filter of
$33'$. In addition, its power spectrum $C_l$ is normalized to that of a CDM 
flat $\Lambda$-model using the HEALPix package. For each of the
simulations the wavelet
coefficients for both the SMHW and the SHW are computed. The SMHW
coefficients are 
computed by convolving the CMB map with the SMHW given in
eq. (13). We again use the HEALPix package to perform such
convolution in Fourier space, having previously calculated the
Legendre coefficients of the SMHW at the specified resolution. The SHW
detail coefficients are computed by performing
the linear combinations of 4 pixels as described in section 3. Computation time
of wavelet coefficients using HEALPix scale as $N^3_{N_{side}}$ and 
$N^2_{N_{side}}$ for SMHW and SHW respectively.

\begin{figure*}
\epsfxsize=14cm
\begin{minipage}{\epsfxsize}\epsffile{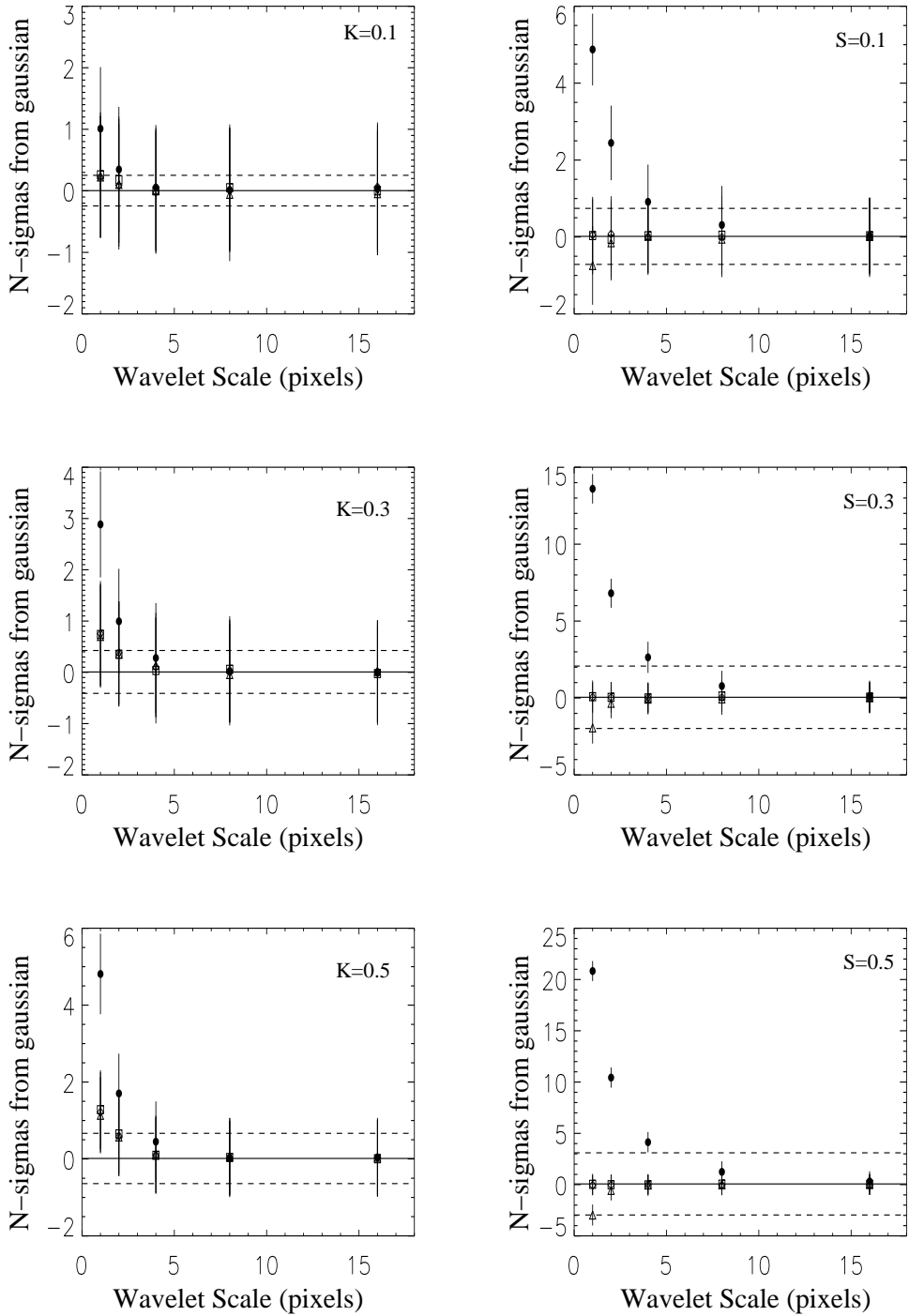}\end{minipage}
\caption{Comparison of Mexican Hat wavelet (black circle) and the
Haar Wavelet details: Vertical (diamond), Diagonal (triangle) and
Horizontal (square); for Kurtosis (left) and Skewness (right) values
of 0.1, 0.3 and 0.5 
(from top to bottom).  Each point represents the number of sigmas
deviated from the Gaussian model. Also plotted is the stripe for the
non-Gaussianity determined from the temperature map (in this case only 
the pixel scale is meaninful, the stripe is drawn only for
illustrative purposes).} 
\label{Sigmas}
\end{figure*}

In figure 4 we show the mean values and dispersion of the skewness
and kurtosis of the Gaussian and non-Gaussian models for the
temperature map, and for the first $5$ resolution levels of the SHW
diagonal, vertical and horizontal coefficients
and SMHW coefficients. As expected the differences are best seen in
the finer resolutions. It is clear from figure 4 that the differences
in the skewness for the two models are more remarkable for the SMHW
than for the SHW and the temperature map. This is also the case for
the kurtosis. 
As we pointed out in section 4.2, there is a strong bias in the
kurtosis of the three details of the SHW coefficients due to the
slight non-uniform distribution of pixels on the sphere in the
HEALPix pixelisation. This kind of bias is expected for any pixelisation of 
the sphere due to the impossibility of having a uniform pixelisation. The 
specific bias introduced will depend on the pixelisation scheme used. 
On the contrary, no bias is present for the SMHW
coefficients due to its continuous nature.    
\begin{figure*}
\epsfxsize=14cm
\begin{minipage}{\epsfxsize}\epsffile{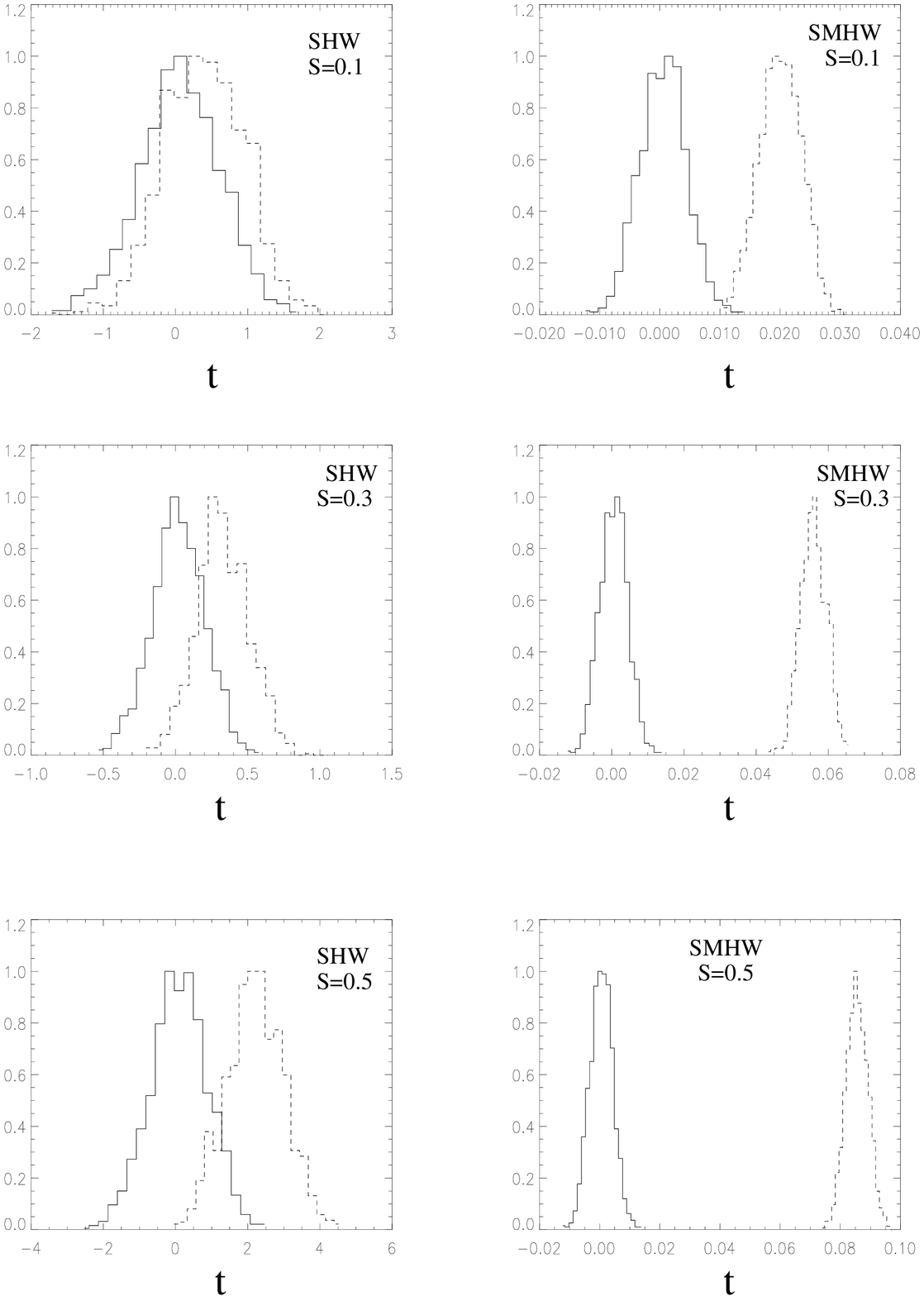}\end{minipage}
\caption{Fisher Discriminant for Skewness from Spherical Haar Wavelet (left) and Spherical Mexican Hat Wavelet (right). From top to bottom, the values of skewness in the original maps are 0.1, 0.3 and 0.5. The solid line is the Gaussian model, while the dashed one represents the non-Gaussian case.} 
\label{SkewPdf}
\end{figure*}
\begin{figure*}
\epsfxsize=14cm
\begin{minipage}{\epsfxsize}\epsffile{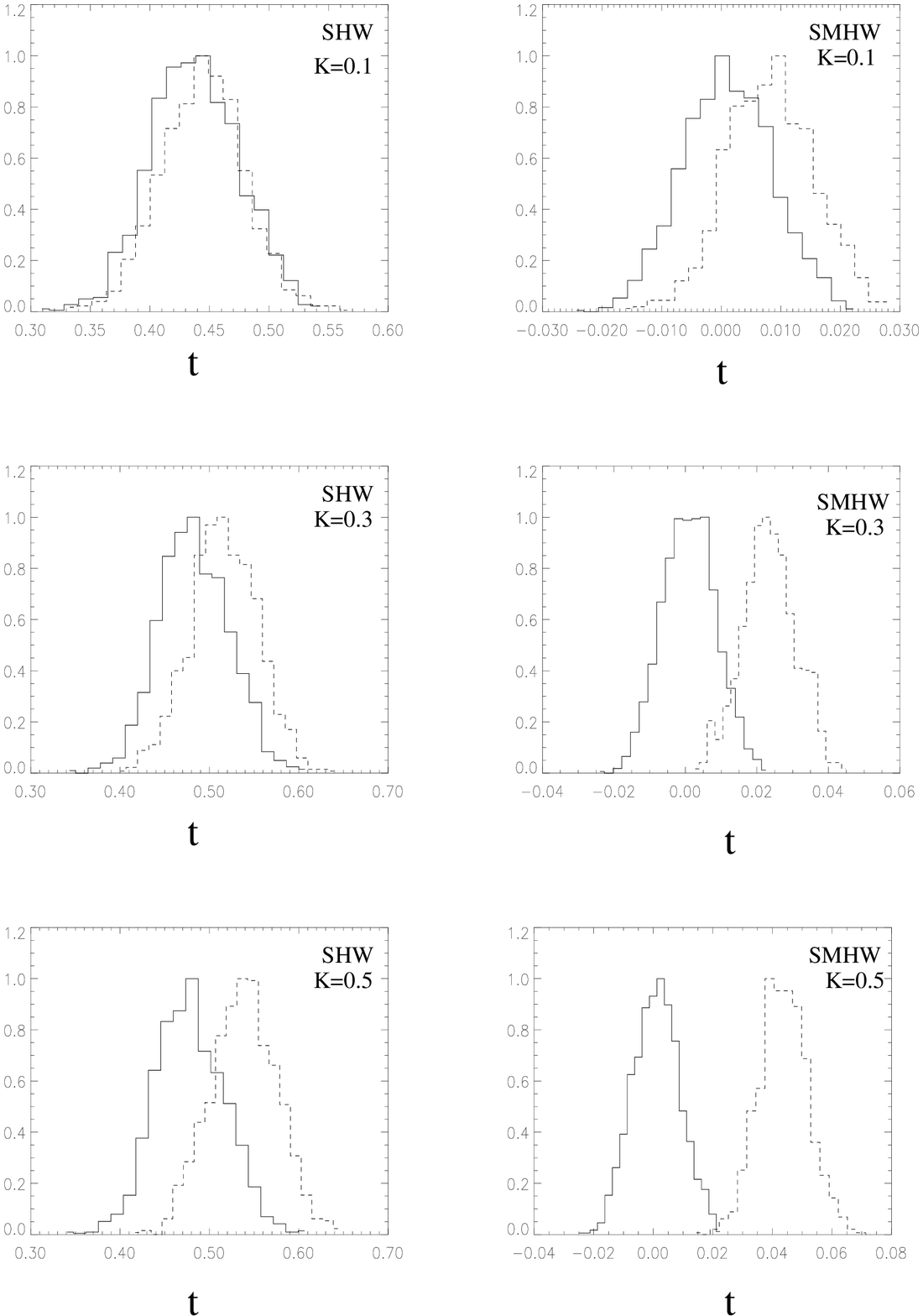}\end{minipage}
\caption{Fisher Discriminant for Kurtosis from Spherical Haar Wavelet
(left) and Spherical Mexican Hat Wavelet (right). From top to bottom,
the values of kurtosis in the original maps are 0.1, 0.3 and 0.5. The
solid line is the Gaussian model, while the dashed one represents the
non-Gaussian case. Please note the bias in the distribution of the SHW 
kurtosis as discussed in the text.} 
\label{KurtPdf}
\end{figure*}
The Fisher discriminant $t$ can still be applied to distinguish between
the two models
even in the presence of that bias in the kurtosis. As seen in the
previous section, what enters in the linear coefficients to compute
the statistic $t$ is the difference
between the means from the two models, cancelling out the bias
term. In figures 5,6 we show the
pdf´s of the statistic $t$ for three values of the skewness and
kurtosis of the non-Gaussian models. It is clear that for both
non-Gaussian models, with either positive skewness or kurtosis, the
SMHW is able to distinguish between the Gaussian and non-Gaussian
models much better than the SHW. 

In table \ref{Fisher} the power $p$ of the Fisher discriminant constructed from
the skewness or kurtosis of the SMHW, SHW and temperature is given for
several values of the cumulants. For the case of the temperature of
the map the statistic is given directly by its
cumulants. Again, the performance of the SMHW is superior to the SHW
and the temperature in all cases.
\begin{figure*}
\epsfxsize=14cm
\begin{minipage}{\epsfxsize}\epsffile{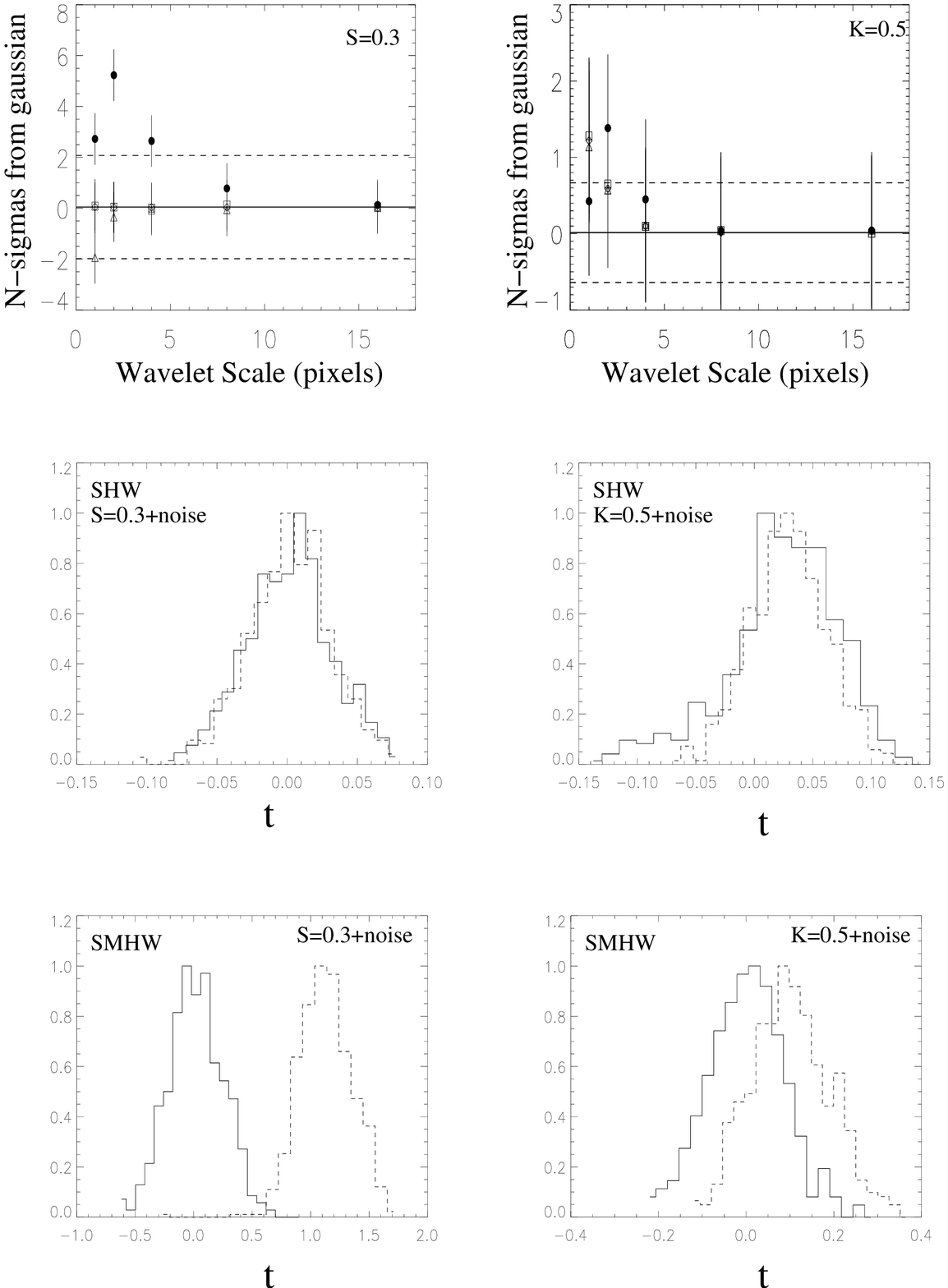}\end{minipage}
\caption{Top graphics: Comparison of Mexican Hat wavelet (black circle) and 
the
Haar Wavelet details: vertical (diamond), diagonal (triangle) and
horizontal (square); for skewness 0.3 (right) and kurtosis 0.5 (left) with 
added noise.  Each point represents the number of sigmas
deviated from the Gaussian model with noise. Also plotted is the stripe for the
non-Gaussianity determined from the temperature map (in this case only 
the pixel scale is meaninful, the stripe is drawn only for
illustrative purposes).
Bottom graphics: Fisher discriminant for skewness (right) and kurtosis (left)
from Spherical Haar and  Mexican Hat Wavelets} 
\label{Noise}
\end{figure*}

Since the SHW is affected by the non-uniform 
pixelisation of the sphere, one might wonder if its failure to detect 
non-Gaussianity is a feature of the Haar wavelet in general or a consequence 
of the 
pixel dependent scale mixing. In order to answer this question we have made 
the same comparison between Gaussian and non-Gaussian models, one with 
skewness 0.3 and the other with kurtosis 0.3, but now on the 
plane. (We have considered simulated $12^\odot.8\times12^\odot.8$ maps with 
$1'.5$ pixels and a beam of $5'$ FWHM. The steps of the simulation and analysis
are the same as for the sphere). The result is very similar to the one found 
on the sphere. Therefore, the failure of the Haar wavelet to detect 
non-Gaussianity is an intrinsic caracteristic of this wavelet and not a 
consequence of the pixel dependent scale 
mixing due to its implementation on the sphere. (Notice, 
however, that its performance can be similar to other planar wavelets for some 
specific features more adapted to its shape, e.g. cosmic strings, see 
Barreiro and Hobson 2001). The pixel 
dependent scale mixing basically 
induces a bias which has been taken into account in the analysis.

In order to know the effect of instrumental noise (white) on the 
discriminating power of the spherical wavelets we have also added
noise to the temperature  maps with an amplitude equals to the signal,
$S/N=1$. In this case 500 simulations were generated. As shown in
figure 7, the first resolution scale is the most affected and now the second 
scale is  the most relevant for discrimination between models. In this 
figure it is also plotted  the new pdf's for the Fisher discriminant
for $30\%$  injected skewness and $50\%$ injected
kurtosis. The noise effect is shown in the narrowing of the separation between
distributions as compared to the no-noise case. We see that the SMHW is still 
able to discriminate with a high 
power for the skewness model with a skewness value in the analysed map of 
$1.1\%$. For the kurtosis model, the addition of noise 
with the same amplitude than the signal
reduces the level of kurtosis in the analysed map from $1.4\%$ to $0.2\%$, 
a level too low to be detectable.

Finally, even if future experiments like MAP and Planck observe the full sky 
probably only the fraction outside the Galactic plane will be used to test 
non-Gaussianity. This problem has already been considered in 
previous papers analysing the Gaussianity of the COBE-DMR data with the SHW and
the SMHW (Barreiro et al. 2000, Cay\'on et al. 2001). As can be seen from 
those papers the impact on the two methods is similar (in both cases one 
looses all the coefficients computed from pixels intersecting the cut).
In any case, for future missions like MAP or Planck, the Galactic cut should be
much smaller than for COBE because of the much better resolution and the much
larger frecuency information, implying a smaller impact on the analysis.

\vspace{1.cm}
\section{Conclusions}

We have compared the performance of the two spherical wavelet families
already used to test the Gaussianity of the COBE-DMR CMB data: Mexican 
Hat (Cay\'on et al. 2001) and Haar (Barreiro et
al. 2000).  As testbed we use non-Gaussian simulations of all-sky arcminute 
resolution CMB maps,
with a power spectrum $C_l$ consistent with observations and artificially 
specified amounts of skewness or kurtosis. Most, if not all,
physically motivated non-Gaussian primordial models of structure
formation proposed in the literature
show some amounts of either of these two moments in the CMB maps. 
These simulated sky maps are pixelised using the widely used HEALPix package.  
As commented in section 4.2 any pixelisation scheme of the sphere will 
introduce a bias because of the impossibility of a uniform pixelisation.
In particular, for the HEALPix scheme 
this bias shows up as a positive kurtosis in the Spherical Haar Wavelets (SHW) 
coefficients even for temperature realizations derived from normal 
distributions. The bias represents a $\approx 10\sigma$ effect for the finest 
resolution, as can be seen from the first row of the kurtosis in table 1. 
On the contrary no bias is present in the case of the Spherical Mexican Hat 
Wavelet (SMHW) due to its continues nature, i.e. not adapted to the 
pixelisation scheme.

The main conclusion of this paper is that the  
SMHW bases are much more efficient to discriminate between Gaussian
and non-Gaussian models with either skewness or kurtosis
present in the CMB maps than the Spherical Haar Wavelet (SHW)
ones. More specifically, the SMHW is able to discriminate a $1.6\%$ skewness
with a power of $100\%$ at the $1\%$ significance level whereas the SHW
can weakly discriminate a $6.9\%$ skewness with a power of only $78\%$ at the 
same significance. In the case of kurtosis, the SMHW detects a $1.4\%$ level
with a power of $99.9\%$ whereas for the SHW the power is only $28\%$, 
at the same $1\%$ significance level. The failure of the Haar wavelet
to detect non-Gaussianity is not a consequence of the pixel dependent scale
mixing due to its implementation on the sphere but an intrinsic 
property of this wavelet (as has been demonstrated by performing a 
similar analysis on the plane). If we were to use the temperature
map instead of the wavelet coefficients, the power would be always
smaller than for the wavelets (only comparable to the SHW in the case
of skewness). An interesting property of the SMHW is that an
injected skewness/kurtosis in the temperature maps produces an
amplified skewness/kurtosis in the SMHW coefficients and a negligible
kurtosis/skewness. On the contrary the SHW is not able to amplify any injected 
skewness/kurtosis with neither skewness nor kurtosis of its coefficients.

Finally, we have also tested the performance of the spherical wavelets in the
more realistic case in which instrumental noise (white) is
present in the maps. In this case the highest resolution scale is the most 
affected, being the best scale for discrimination the second one. For a 
signal to noise ratio $S/N=1$, and combining all the information from
all the scales with the Fisher discriminant, the SMHW is still
able to discriminate with a high power levels of skewness and kurtosis above 
$1\%$.

\section*{Acknowledgments}

We thank R. Bel\'en Barreiro and K. Gorski for helpful comments. We
acknowledge partial financial support from the Spanish CICYT-European
Commission FEDER project 1FD97-1769-C04-01, Spanish DGSIC project
PB98-0531-C02-01 and INTAS project INTAS-OPEN-97-1192.

This work has used the software package HEALPix (Hierarchical, Equal
Area and iso-Latitude Pixelisation of the
sphere, http://www.eso.org/science/healpix), developed by
K.M. Gorski, E.F. Hivon, B.D. Wandelt, J. Banday, F.K. Hansen and M. 
Barthelmann.


\begin{thebibliography}{}

\bibitem{}Aghanim, N., Forni, O. \& Bouchet, F.R. 2000, astro-ph/0009463

\bibitem{}Antoine, J.-P. \& Vandergheynst, P., 1998, {\it J. Math. Phys.}, 39, 3987



\bibitem{}Barreiro, R.B. \& Hobson, M.P. 2001, MNRAS, 327, 813 
(astro-ph/0104300) 



\bibitem{}Barreiro, R.B., Mart\'\i nez-Gonz\'alez, E. \& Sanz,
J.L. 2001, MNRAS, 322, 411

\bibitem{}Barreiro, R.B., Hobson, M.P., Lasenby, A.N, Banday, A.J., G\'{o}rski, K.M. \& Hinshaw, G. 2000, MNRAS, 318, 475





\bibitem{}Bromley, B.C., Tegmark, M. 1999, ApJ, 524, L79

\bibitem{}Cay\'on, L., Sanz, J. L., Barreiro, R. B., Vielva, P., Toffolatti,
    L., Silk, J., Diego, J. M. \& Argueso, F., 2000, MNRAS, 315, 757  

\bibitem{} Cay\'on, L., Sanz, J.L., Mart\'\i nez-Gonz\'alez, E., Banday, A.J.,
Arg$\ddot{u}$eso, F., Gallegos, J.E., Gorski, K.M. \& Hinshaw,
G. 2001, {\it MNRAS}, 2001, 326, 1243.

\bibitem{}Contaldi, C.R., Ferreira, P.G., Magueijo, J., G\'orski,
K.M. 2000, ApJ, 534, 25

\bibitem{}Coulson, D., Ferreira, P., Graham, P., Turok, N. 1994,
Nature, 368, 27 

\bibitem{}Cowan, G., 1998, ``Statistical Data Analysis'', Oxford
University Press, Oxford

\bibitem{}Diego, J.M., Mart\'\i nez-Gonz\'alez, E., Sanz, J.L.,
Mollerach, S. \& Mart\'\i nez, V.J. 1999, MNRAS, 306, 427

\bibitem{}Ferreira, P.J., Magueijo, J. \& G\'{o}rski, K. 1998, ApJ, 503, L1 

\bibitem{}Fisher R.A., 1936, Ann. Eugen., 7, 179; reprinted in
``Contributions to Mathematical Statistics'', 1950, Jonh Wiley, New York


\bibitem{}G\'orski, K.M., Hivon, E. \& Wandelt, B.D. (astro-ph/9812350) 1999, Proceedings of the MPA/ESO Conference on Evolution of Large-Scale Structure: from Recombination to Garching, 2-7 August 1998; eds. A.J. Banday, R.K. Sheth and
L. Da Costa, PrintPartners IPSKAMP NL (1999)



\bibitem{}Heavens, A.F., Gupta, S. 2001, MNRAS, 324, 960

\bibitem{}Hobson, M.P., Jones, A.W. \& Lasenby, A.N. 1999, MNRAS, 309, 125

\bibitem{}Kaiser, N., Stebbins, A. 1984, Nature, 310, 391


\bibitem{}

\bibitem{}Komatsu, E. \& Spergel, D.N. 2001, PRD, 63, 63002

\bibitem{}Linde, A. \& Mukhanov, V. 1997, PRD, 56, 535




\bibitem{}Magueijo, J. 2000, ApJ, 528, L57

\bibitem{} Marr, D. \& Hildreth, E.C. 1980, {\it Proc. Roy. Soc. London, Ser. B}, No. 207, 187.


\bibitem{}Mollerach, S., Mart\'\i nez, V.J., Diego, J.M., 
Mart\'\i nez-Gonz\'alez, E., Sanz, J.L. \& Paredes, S. 1999, ApJ, 525, 17

\bibitem{}Mukherjee, P., Hobson, M.P. \& Lasenby, A.N. 2000, MNRAS, 318, 1157

\bibitem{}Naselsky, P.D., Novikov, D.I. 1998, ApJ, 507, 31

\bibitem{}Netterfield, C.B. et al. 2001, astro-ph/0104460


\bibitem{}Pando, J. Valls-Gabaud, D., Fang, L.Z. 1998, Phys.Rev.Lett., 81, 4568

\bibitem{}Pryke, C. et al. 2001, astro-ph/0104490


\bibitem{}Rocha, G. et al. 2000, astro-ph/0008070

\bibitem{}Schmalzing, J. \& G\'{o}rski, K.M. 1998, MNRAS, 297, 355

\bibitem{}Stompor, R. et al. 2001, astro-ph/0105062

\bibitem{}Sweldens, W., 1996, Applied Comput. Harm. Anal., 3, 1186

\bibitem{}Tenorio, L., Jaffe, A.H., Hanany, S. \& Lineweaver,
C.H. 1999, MNRAS, 310, 823 


\bibitem{}Turok, N, Spergel, D. 1990, PRL, 64, 2736

\bibitem{}Verde, L., Wang, L., Heavens, A.F. \& Kamionkowski, M. 2000, 
MNRAS, 313, 141

\bibitem{}Vielva, P., Mart\'\i nez-Gonz\'alez, E., Cay\'on, L., Diego, 
J.M., Sanz, J.L., Toffolatti, L. 2001, MNRAS, 326, 181



\bibitem{}Wu, J.H.P. et al. 2001, astro-ph/0104248


\end{thebibliography}
\end{document}

\end